\documentclass[prd,showkeys,preprint,11pt]{revtex4}
\usepackage{graphicx}
\usepackage[breaklinks=true,colorlinks=true,linkcolor=blue,urlcolor=blue,citecolor=blue]{hyperref}
\usepackage[british]{babel}
\RequirePackage{flushend}
\usepackage{amsmath,amsfonts,amssymb}
 \expandafter\let\csname equation*\endcsname\relax
  \expandafter\let\csname endequation*\endcsname\relax
\usepackage[mathcal]{eucal}
\usepackage{mathrsfs}
\usepackage{latexsym}
\usepackage{bm}
\usepackage[all]{xy}
\usepackage{dsfont}
\usepackage{upgreek}
\newcommand{\sech}{\text{sech}}
\newcommand{\ex}{\text{e}}
\newcommand{\upd}{\text{d}}

\begin{document}

\title{Exotic Elko on string-like defects in six dimensions}

\author{D. M. Dantas}
\email{davi@fisica.ufc.br}
\affiliation{Universidade Federal do Cear\'a (UFC), Departamento de F\'{\i}sica, Campus do Pici, Caixa Postal 6030, 60455-760, Fortaleza, Cear\'{a}, Brazil}


\author{Rold\~{a}o da Rocha}
\email{roldao.rocha@ufabc.edu.br}
\affiliation{Centro de Matem\'atica, Computa\c c\~ao e Cogni\c c\~ao, Universidade Federal do ABC, 09210-580, Santo Andr\'{e}
- SP, Brazil}

\author{C. A. S. Almeida}
\email{carlos@fisica.ufc.br}
\affiliation{Universidade Federal do Cear\'a (UFC), Departamento de F\'{\i}sica, Campus do Pici, Caixa Postal 6030, 60455-760, Fortaleza, Cear\'{a}, Brazil}




%


\begin{abstract}
We analyse the trapping of eigenspinors of the charge conjugation operator with dual helicity (Elko), in thin and thick string-like models with codimension-2.  Elko spinor fields describe mass dimension one fermions in four dimensions (and, correspondingly, mass dimension two fermions in six dimensions), that represent natural dark matter prime candidates.  This dark spinor has many applications, from particle physics to cosmology.  On the other hand, six-dimensional brane-world models have, among other prominent features,  the spontaneous confinement of free spin 1 fields and a mechanism that explains the mass hierarchy of fundamental fermions.  In this paper,  we use scalar couplings in order to confine the zero mode of Elko in six dimensions.  Moreover, we use the Elko dark spinor features to propose an exotic coupling in order to remove the complex-valued terms in the massive Kaluza-Klein modes. Hence, we show that six dimensional models can resolve the main issues of Elko fields confinement presented in five dimensions.
\end{abstract}

\maketitle

\section{Introduction}
Warped brane-world scenarios, like the Randall-Sundrum (RS)  models  \cite{RS1, RS2}, proposed that our world is constituted by a three-brane embedded in a warped higher dimensional  space-time. With this proposition, one can solve the hierarchy problem and explain several issues in various branches of Physics \cite{RS1,RS2,cou1, Davi5,R8, Davi4, Davi7,valle,condu,lhc2}. Focusing on the recent publications, brane-worlds models were used to: perform bounds into corrections to Coulomb's law \cite{cou1, Davi5} and to set limits in the  electrical conductivity \cite{condu}, studied by the informational entropy \cite{R8, Davi4, Davi7}, to respond to anomalies in the meson $B$ decay \cite{lhc2}, and to explain issues in the neutrinos physics \cite{valle}, for instance. 

This paper focus on six-dimensional (6D) anti-de Sitter ($AdS_6$) brane-worlds, which have some interesting features. In fact, the mass hierarchy is solved without any requirement of fine tuning between the bulk cosmological constant and the brane tension \cite{GS1}, with  correction $\mathcal{O}(d^{-3})$ to the Newtonian potential \cite{GS1, Silva:2012yj, Diego, Davi3, Davi6} that is smaller than within five-dimensional (5D) models \cite{RS2}. Moreover, 6D models support the localisation of free gauge zero modes,  even in the thin brane case \cite{Oda1, Coni2, Torre}, whereas  a scalar coupling in the $AdS_5$ case is demanded \cite{Davi5}. In Lorentz violation scenarios, a  massless four-dimensional (4D) graviton can be confined in 6D \cite{victor6d}, but it is not possible in the thin 5D model \cite{Rizzo}. Fermions fields in 6D have some prominent applications \cite{Liu1}, and Ref. \cite{Merab} suggests defects with two angular extra dimensions, where the angular momentum in the transverse space of trapped three fermion zero modes   is correlated to the three generations of fundamental fermions in 4D. Additionally, Ref. \cite{Frere} uses a single fermion family in 6D to explain the mass hierarchy of neutrinos.

 On other hand, Elko  spinor fields (dual-helicity eigenspinors of the charge conjugation operator) \cite{chengk, LiuEK, marcao, allu1,Ahluwalia:2015vea,Fabbri:2010ws}
 are  spin-1/2  matter fields, with features that put such fermionic matter fields as prime candidates to describe dark matter \cite{Agarwal,Ahluwalia:2010zn,Fabbri:2010va,Ahluwalia:2009rh,Ahluwalia:2008xi,Boehmer:2006qq}. Such spinor fields have mass dimension one (in 4D), hence Elko interactions with the standard model  matter and gauge fields are suppressed by at least one order of magnitude of unification scale. This means that the interactions of Elko spinor fields are merely limited  to gravitons and the Higgs field. The Hawking radiation and further properties, regarding this matter field, have been explored \cite{daRocha:2014dla}. From the phenomenological point of view, Elko can be produced by Higgs interactions  \cite{marcao}, and realise a particle, whose symmetries are governed by the Very Special Relativity (VSR) \cite{VSR}. Some attempts to detect Elko at the LHC have been moreover proposed \cite{marcao}, as well as promising  applications in cosmology \cite{saulo, bur1}. 
 
 The Elko spinor has been localised in 5D brane-worlds  \cite{LiuEK,BroEK1}, where has been verified that the zero mode trapping  requires scalar couplings. Hence, as usual for spinors fields in 5D \cite{t1}, a Yukawa-type interaction between Elko spinor fields and scalar fields is used in order to confine the massless mode. In Ref. \cite{LiuEK}, a scalar field with mass value was used for the thin 5D model and the scalar kink field for the thick 5D model. Moreover, Ref. \cite{BroEK1} proposed a geometric coupling with the Ricci scalar. However, in both cases, the massive modes present complex values, which maed it impossible to compute resonances \cite{LiuEK, BroEK1}. We demonstrate in this paper that the complex-valued  Elko massive eigenfunctions can be removed by a covariant derivative specific coupling.  In fact, regarding spin $1/2$ Dirac (or Weyl) fermions and spin $3/2$ Rarita-Schwinger fermions in 6D models, minimal couplings with an electromagnetic vector $U(1)$ gauge background field are performed in order to bound the zero mode \cite{Liu1, Davi1, Davi2, Davi3}. However, since Elko is a dark spinor, there is no interaction with electromagnetic gauge field, but an exotic spin structure term (such as 1-form field representing an element of the cohomology group $H^1(M, \mathbb{Z}_2)$) is allowed though \cite{exotic, alex,isham0,petry,avis0}. These exotic spinor structures play a prominent role in the Elko models framework, which  no  another mass dimension 3/2 matter field can manifest.  In that regard, a mass generation mechanism for the  mass dimension one spinors  has been derived, by coupling with a kink, in the context of a $\lambda \phi^{4}$ field theory \cite{alex}. 
\par 
 In the present letter, we prove that in 6D brane-worlds the Elko  zero mode is only confined with a scalar coupling term. Moreover, in order to remove the Elko massive complex-valued eigenfunction, we introduce a 4D exotic coupling of Elko spinor fields \cite{exotic, alex}, which also prevents the existence of bounded massive modes. To this end, we made a brief review of string-like defects, using the Gergheta-Shaposhnikov (GS) thin model \cite{GS1, Oda1, Liu1} and the Hamilton string-cigar thick model (HC) \cite{Davi3, Silva:2012yj,Diego}. Next, we show the task of the Elko spinor confinement in string-like models. Further, Elko modes issues are scrutinised in the GS model, where the calculus can be achieved analytically. We present the values of scalar fields and coupling constant in order to trap the Elko zero mode and propose a topological exotic term which allows the proper treatment of massive Kaluza-Klein (KK) modes. Hence, we point the numerical values for couplings and  present the generalisation to localise the Elko spinor on any 6D string-like model. Finally,   our main results and perspectives are summarised in the conclusions.
\section{The Elko matter fields in six dimensions}{
The metric ansatz for 6D string-like models can be represented by \cite{GS1, Oda1}
\begin{gather}
\upd s^2_6=F(r)\eta_{\mu \nu}\upd x^{\mu}\upd x^{\nu}+\upd r^2+H(r)\upd \theta^2 ,
\label{6dmetric}
\end{gather}
We consider for this ansatz the signature for the $\mathcal{M}_4$ metric as $\eta_{\mu \nu}= diag(-1,+1,+1,+1)$. The warp factors $F$ and $H$ do solely  depend upon the radial coordinate $r$, that is restricted to $r  \in \left[0, \infty\right)$, whereas the angular coordinate ranges   $\theta \in \left[0,2\pi\right)$.
\par 
The Ricci scalar is given by:
\begin{gather}
R=-\left[4\frac{ F''}{F}+2\frac{ F'}{F}\frac{H'}{ H}+\left(\frac{F'}{F}\right)^2+\frac{H''}{H}-\frac{1}{2}\left(\frac{H'}{H}\right)^2\right]\ .
\label{sricci}
\end{gather}
\par 
Based on the above mentioned framework, we present the first 6D model proposed by Gergheta-Shaposhnikov and so-called \textit{string-like defect} (GS) \cite{GS1,Oda1, Liu1}. For the vacuum solution  and positive string tension such model provides the following metric coefficients  \cite{GS1,Oda1,Liu1}:
\begin{gather}\label{gs-string}
F_{GS}(r)=\ex^{-cr}, \quad H_{GS}(r)=R^2_0 F_{GS}(r)\,,
\end{gather}
where the parameters $c$ and $R_0$ are positive constants, where $c^2=-\frac{2}{5} \frac{\Lambda}{M^4_6}$ and $R_0$ is an arbitrary  length scale constant \cite{GS1,Oda1, Liu1}. The GS model configures a thin $AdS_6$ space with curvature obtained by application of Eq. \eqref{gs-string} in Eq. \eqref{sricci} given by:
\begin{equation}\label{gs-ricci}
R_{GS}=-\frac{15}{2}c^2\ .
\end{equation}

In order to solve some issues involving the regularity and energy conditions, we use in this paper the regular thick model so-called \textit{Hamilton string-cigar} (HC)\cite{Silva:2012yj, Diego, Davi3}:
\begin{equation}\label{hc-string}
F_{HC}(r)=\ex^{-cr+\tanh(cr)}, \quad H_{HC}(r)=\left(\frac{\tanh{cr}}{c}\right)^2F_{HC}(r) .
\end{equation}
This geometry is built upon a warped product between a 3-brane and the cigar soliton space, which is a 2D stationary solution for the Ricci flow \cite{Silva:2012yj,Diego, Davi3}. Moreover, the changes performed by this regular HC model in comparison to the non-regular GS model arise just close to the origin. Hence, asymptotically Eq. \eqref{gs-string} is retrieved. Fig. \ref{curv-ricci} shows that the curvature of the HC model converges asymptotically to the GS curvature \eqref{gs-ricci}.
\begin{figure}[!htb]
       \centering
                \includegraphics[width=.68\linewidth]{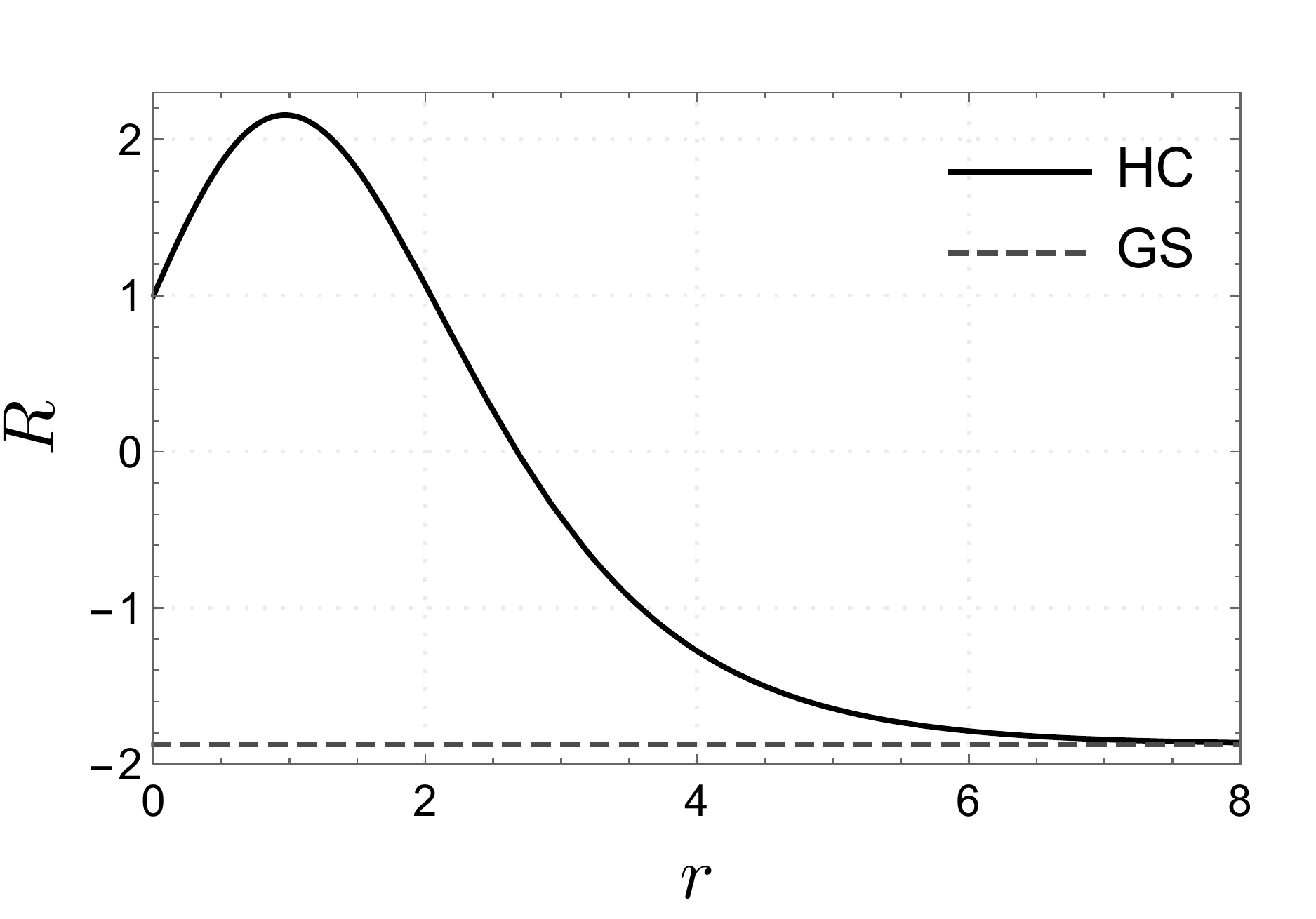}
                \caption{Ricci curvature scalar for the HC model (black line) and the GS model (gray dashed line), both for $c=0.5$.}
                \label{curv-ricci}
\end{figure}

Now, we study the trapping of Elko fermions, where we shall  conclude that there is no possibility of confining the zero mode of Elko fields in 6D models without couplings terms. Based upon the Elko 5D action of Ref. \cite{LiuEK}, the action for bulk massless Elko spinors in six dimensions is given by
\begin{equation}\label{action}
\!\!\!S\!=\!-\!\int{\upd x^6}\!\sqrt{-g}\left[\frac{g^{MN}}{4}\left(D_{M}\bar{\Upsilon}D_{N}\Upsilon+D_{N}\bar{\Upsilon}D_{M}\Upsilon\right)+ \zeta_1\bar{\Upsilon}\mathcal{F}_1\Upsilon \right].
\end{equation}
where $D_M$ is the covariant derivative and $\Upsilon$ represents the $8\times1$ mass dimension two Elko spinor field,  the $\mathcal{F}_1(r)$ is a mass dimension two scalar field  and $\zeta_1$ a dimensionless coupling constant.  The following equation of motion is obtained:
\begin{gather}
\left[\frac{1}{\sqrt{-g}}D_M\left(\sqrt{-g}g^{MN}D_N\right)-2 \zeta_1\mathcal{F}_1\right]\Upsilon \ =0 \ ,
\label{eqm}
\end{gather}
Explicitly,  in the absence of an exotic term,  the covariant derivative  obtained from \eqref{6dmetric} reads
\begin{gather}
D_M \Upsilon=\left[\partial_M+ \Omega_M\right]\Upsilon,
\label{dcov}
\end{gather}
the non-vanishing terms of spin connections take the forms $\Omega_{\mu}(r)=\frac{1}{4}\frac{F^{\prime}}{\sqrt{F}}\Gamma_{\bar{\mu}}\Gamma_{\bar{r}}$ and $\Omega_{\theta}(r)=\frac{1}{4}\frac{H^{\prime}}{\sqrt{H}}\Gamma_{\bar{\theta}}\Gamma_{\bar{r}}$, where
the primes denote the derivative with respect to $r$, whereas $\Gamma_{\bar{M}}$ are the 6D gamma matrices  in the flat space, which can be represented in terms of 4D flat gamma matrices $\gamma_{\bar{M}}$ \cite{Oda1, Liu1, Davi3}. We choose the usual Weyl decomposition for spinors in 6D models, the 6D gamma matrices and the Kaluza-Klein spinor decomposition \cite{Oda1, Liu1, Davi3}
\begin{gather}
\label{rowspinor}
\!\!\!\Upsilon(x,r,\theta)=
\begin{pmatrix}
\lambda\\
0\\
\end{pmatrix}, \  \lambda(x^{\mu},r,\theta)=\sum\limits_{n,l}\lambda_n(x^\mu)\varepsilon_n(r)\lambda_{l}(\theta),
\end{gather}
where $\lambda_{n}(x)=\varsigma^{(n)}_{\pm}(x^{\mu})+\tau^{(n)}_{\pm}(x)$ represents  the four types of mass dimension one 4D Elko \cite{LiuEK,BroEK1}. The  $\varepsilon_n(r)$ is the radial component which must be confined, whereas the $\lambda_{l}(\theta)=\ex^{il\theta}$ denotes the angular component with $l$ the orbital number \cite{GS1, Oda1, Liu1}. The index $n$ labels the values of masses $m_n$. It is worth to mention that this spinor is indeed an eigenspinor of the 6D charge conjugation operator, and restricted to 4D is then an eigenspinor of the corresponding 4D charge conjugation operator. The result is interesting and it means that
the KK modes of different types of an Elko spinor are
the same and indistinguishable \cite{LiuEK}.

According to this decomposition, the 4D flat gamma matrices act upon  4D  spinors $\lambda_{n}(x)$ as \cite{LiuEK, BroEK1}:
\begin{gather}
\gamma^{\mu}\partial_{\mu}\varsigma_{\pm}(x)=\mp im\varsigma_{\mp}(x); \quad \gamma^{\mu}\partial_{\mu}\tau_{\pm}(x)=\pm im \tau_{\mp}(x);\nonumber \\
\gamma^{r}\varsigma_{\pm}(x)=\pm \tau_{\mp}(x); \ \quad \gamma^r\tau_{\pm}(x)=\mp \varsigma_{\mp}(x); \nonumber \\
\gamma^{\theta}\varsigma_{\pm}(x)=\mp i\varsigma_{\mp}(x); \ \qquad \gamma^{\theta}\tau_{\pm}(x)=\pm i\tau_{\mp}(x). \nonumber
\end{gather}
It is worth to emphasize that Elko spinors do not obey a Dirac equation, since the Dirac spinor $\psi$ satisfies $i\gamma^{\mu}\partial_{\mu}\psi=m\psi$ \cite{LiuEK,BroEK1}.

Turning to Eq. \eqref{eqm} the 4D section yields
\begin{gather}
\left[\frac{D_{\mu}}{\sqrt{-g}}\left(\frac{\sqrt{-g}\eta^{\mu \nu}}{F(r)}D_{\nu}\right)\right]\Upsilon= \ \nonumber \\
 \sum\limits_{n,l}\lambda_n(x)\lambda_{l}(\theta)\frac{1}{F}\left[m_n^2-\frac{im_nF'}{2\sqrt{F}}-\frac{1}{4}\left(\frac{F'}{\sqrt{F}}\right)^2 \right]\varepsilon_n(r)\,. \label{dmu}
\end{gather} 
Moreover, the radial part yields
\begin{gather}
\left[\frac{\partial_{r}\left(\sqrt{-g}\partial_{r}\right)}{\sqrt{-g}}-2 \zeta_1\mathcal{F}_1\right]\Upsilon= \nonumber \\ \sum\limits_{n,l}\lambda_n(x)\lambda_{l}(\theta)\left[\partial_r^2-2\left(\frac{F'}{F}+\frac{1}{4}\frac{H'}{H}\right)\partial_r-2 \zeta_1\mathcal{F}_1\right]\varepsilon_n(r)\,. \label{drr}
  \end{gather}
Finally, the angular component of the  equation of motion is given by
 \begin{gather}
\left[\frac{D_{\theta}}{\sqrt{-g}}\left(\frac{\sqrt{-g}}{H(r)}D_{\theta}\right)\right]\Upsilon \
=\nonumber \\
 \!\!\!\!\sum\limits_{n,l}\lambda_n(x)\lambda_{l}(\theta)\frac{1}{H}\left[\frac{l}{2}\frac{H^{\prime}}{\sqrt{H}}-l^2-\frac{1}{16}\left(\frac{H^{\prime}}{\sqrt{H}}\right)^2\right]\varepsilon_n(r). \label{dtheta}
 \end{gather}

The sum of Eqs. \eqref{dmu}, \eqref{drr} and \eqref{dtheta}
results in a complete equation of motion \eqref{eqm}, where the annulment condition befalls  in the radial component $\varepsilon_n(r)$. For the $s$-wave solution ($l=0$), we obtain the following second order differential  equation:
 \begin{gather}
\left[\partial_r^2 + \mathcal{P}(r)\partial_r+\mathcal{Q}(r)\right]\varepsilon_n(r)=0
\label{sl}
 \end{gather}
where the coefficients are:
\begin{eqnarray}
\!\!\!\! \!\!\!\! \!\!\!\! \!\!\!\! \mathcal{P}(r)\!\!\!\!&\!=\!\!&\!\!2\frac{F'}{F}+\frac{1}{2}\frac{H'}{H}\ ,
 \\\!\!\!\! \!\!\!\!  \!\!\!\! \!\!\!\! \mathcal{Q}_n(r)\!\!\!\!\!&\!=\!\!\!&\!\!\!\!\frac{m_n^2}{F(r)}\!-\!\frac{im_nF'}{2F^{3/2}}\!-\!\left[\left(\frac{F'}{2F}\right)^{\!\!2}\!\!+\!\left(\frac{H'}{4H}\right)^{\!\!2}\right]\!-\! 2 \zeta_1\mathcal{F}_1.
\label{pq}
 \end{eqnarray}

To consider a solution of Eq. \eqref{sl} as confined mode, it must obey certain conditions. In fact, the boundary conditions due to the axial symmetry \cite{GS1,Oda1,Davi3} read \begin{gather}
\varepsilon_n'(0)= \varepsilon_n'(\infty)=0\,,
\label{normb}
\end{gather}
and the  orthonormality condition
\begin{gather}
\int_0^\infty{ F(r)\sqrt{H(r)} \varepsilon_n^*(r)\varepsilon_s(r) }\upd r=\delta_{ns}
\label{normc}
 \end{gather} must hold as well.
Hereafter we particularise these results for the  string-like brane-worlds models.}

\section{Confining Elko in the Thin model}
We start to analyse the Elko matter fields on the GS thin string-like defect, where the method can be performed analytically. We expose the forms of scalar fields that provide a confined zero mode and real-evaluated massive solutions. The Eq. \eqref{sl} with the warp factors of  GS string in Eq. \eqref{gs-string} yields
\begin{gather}
\varepsilon_{n}''(r)\!-\!\frac{5c}{2}\varepsilon_{{n}}'(r)\!+\!\left(m^2_n\ex^{cr}\!-\!\frac{im_nc}{2}\ex^{\frac{cr}{2}}\!-\!\frac{5c^2}{16}\!-\! 2 \zeta_1\mathcal{F}_1\right)\varepsilon_{{n}}(r)=0. 
\label{sl-gs}
 \end{gather}
 
\begin{figure}[!htb]
       \centering
                \includegraphics[width=.68\linewidth]{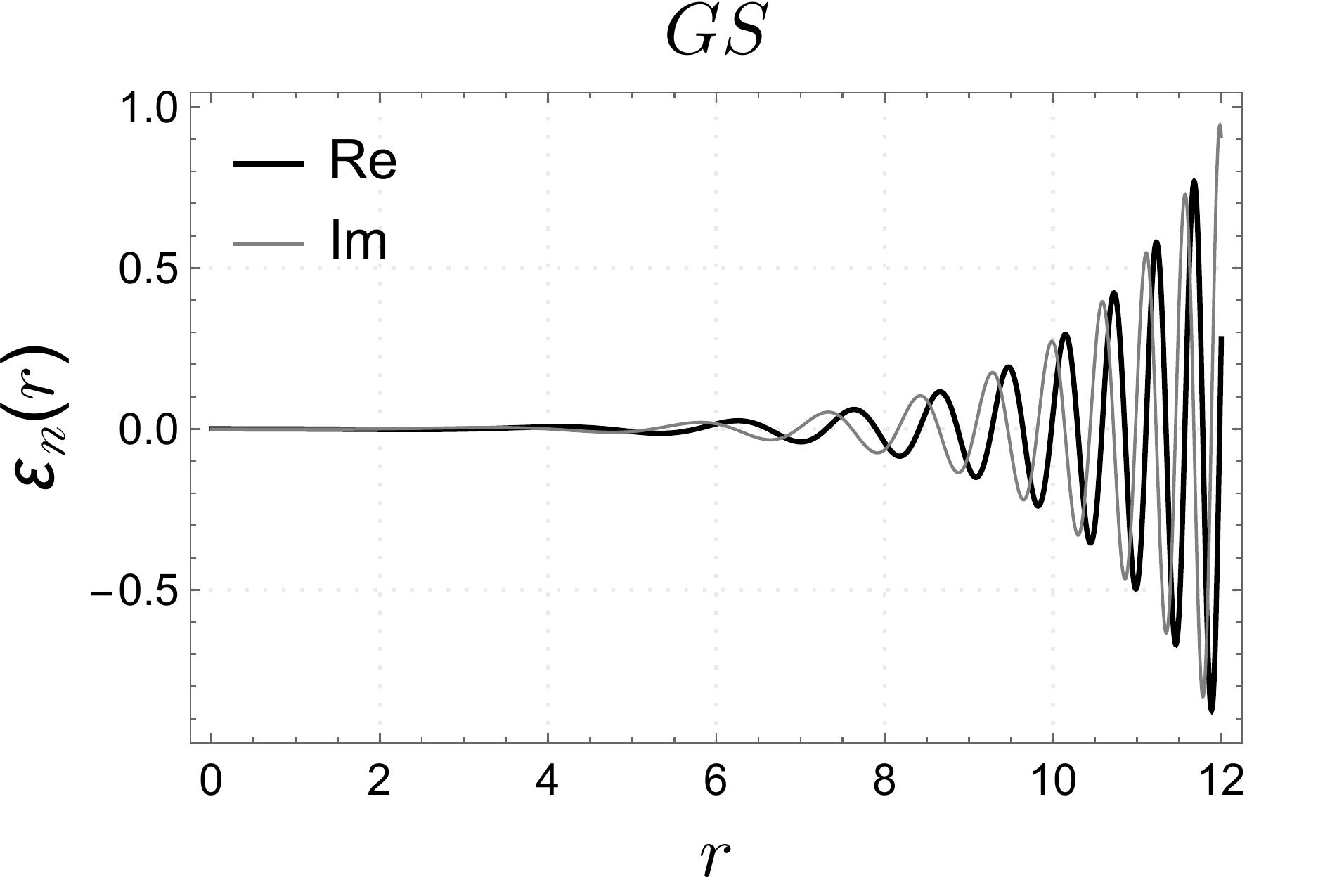}
                \caption{Solutions  in real (thick line) and imaginary (thin line) parts of the eigenfunction for GS string in Eq. \eqref{m-gs}. We set $c=0.5, m_{n}=0.80$, $R_0=1.00$ and the constants $N_1^n=N_2^n=0.05$.}
                \label{GS-mass}
\end{figure}

For the  massless mode  we have
\begin{gather}
\varepsilon_{0}''(r)-\frac{5c}{2}\varepsilon_{{0}}'(r)-\left(\frac{5 c^2}{16}+ 2 \zeta_1\mathcal{F}_1\right)\varepsilon_{{0}}(r)=0.
\label{sl-gs-0}
 \end{gather}
In order to work with constant coefficients differential equations, we assume $\mathcal{F}_1$ constant, as expected for thin  models \cite{LiuEK, BroEK1}, where there is a scalar field with mass value in Ref.  \cite{LiuEK} or the constant curvature term \cite{BroEK1}. Hence, the solution reads
\begin{gather}\label{0-gs}
 \!\!\!\!\varepsilon_{0}(r)=N_1 \ex^{\left(\frac{5 c}{4}- \sqrt{\frac{15}{8}c^2+2\zeta_1\mathcal{F}}\right) r}+N_2 \ex^{\left(\frac{5 c}{4}+ \sqrt{\frac{15}{8}c^2+2\zeta_1\mathcal{F}}\right) r} \ ,  
 \end{gather}
where  $N_1$, $N_2$ are real arbitrary  constants with values obtained by application of  conditions \eqref{normb} and \eqref{normc}. Due to the boundary condition in Eq. \eqref{normb}, we have that the single  value of coupling term $\zeta_1 \mathcal{F}_1$ that gives a non-null solution is
\begin{gather}\label{f-fix}
\zeta_1\mathcal{F}_1=-\frac{5}{32}c^2 \ .
\end{gather}
The physical significance of this term $\mathcal{F}_1$ in the thin model can be interpreted both as a squared mass value \cite{LiuEK, BroEK1} $\left(M_{Elko}^2 = c^2\right)$  with the coupling constant  $\zeta_1=-5/32$. Or else as a geometrical coupling \cite{BroEK1} with  the scalar curvature of the GS model in Eq. \eqref{gs-ricci} with $\zeta_1=1/16$. In fact, the solution \eqref{0-gs} with imposition of Eq. \eqref{f-fix}, that obeys both conditions \eqref{normb} and \eqref{normc}, leads to a constant solution for the Elko zero mode as:
\begin{gather}\label{e-0}
\varepsilon_{0}(r)=\sqrt{\frac{3}{2 R_0}c} \ .
\end{gather}
This is the same solution for the  zero mode of gravity \cite{GS1} and the scalar fields in 6D \cite{Davi3}. For massive modes, Eq. \eqref{sl} in the GS string with the choice in Eq. \eqref{f-fix} presents the following complex-valued solution:
\begin{eqnarray}\label{m-gs}
\varepsilon_n(r)= \ex^{\frac{5c}{4}r}\ex^{-\frac{\mathcal{X}_n}{2}}\left[N_1^n U\left(\frac{5}{2},6,\mathcal{X}_n\right)+ N_2^n L_{\frac{5}{2}}^{5}\left(\mathcal{X}_n\right)\right],
\end{eqnarray}
where  $\mathcal{X}_n(r)=4\frac{i m_n}{c}e^{\frac{c r}{2}}$, the $U$ is the confluent hypergeometric function of the second kind, $L$ is the  generalised Laguerre polynomial, and  $N_1^n, N_2^n$ are normalisation constants. The amplitude term $\ex^{\frac{5c}{4}r}$ in \eqref{m-gs} makes this expression to grow exponentially, whereas the other terms are oscillating. Hence the boundary condition \eqref{normb} is not satisfied at $r\to\infty$ as can be seen in Fig.  \ref{GS-mass}. Even if the boundary condition \eqref{normb} is ignored, the growth of $\lvert \varepsilon_n(r)\lvert^2$ is slowed by $F\sqrt{H}=R_0\ex^{-\frac{3}{4}cr}$ in Eq. \eqref{normc}. Nevertheless, this integrand still increases and can not be normalised. Thus, there is no bounded massive Elko in the GS  6D thin model. It is interesting to point out that in the 5D case, for the thin RS model, Refs. \cite{LiuEK,BroEK1} found Whittaker functions for Elko massive solutions, which are complex-valued as well.
This complex-valued result is generated by the term $-\frac{im_nc}{2}\ex^{\frac{cr}{2}}$ in Eq. \eqref{sl-gs},  which comes from the Elko 4D portion in Eq. \eqref{dmu}.

 A way to prevent this result is to modify the covariant derivative in similar form to proposed in Refs. \cite{alex, exotic}. The covariant derivative acting on exotic spinor fields changes by an additional 1-form field that defines an integer cohomology class in the  $\check{\rm C}$ech sense \cite{isham0,petry,avis0}, encrypting the non-trivial topology.  {The first example of a non-trivial topology is that probed by an electron in the vicinity of a wire with a current, which is the source of the Aharonov-Bohm effect.  
Considering a trivial topology is a quite strict framework  \cite{isham0}. In fact, the path integral setup includes multiply connected spaces  \cite{Hawking2,duff}, also employed in the study of superconductivity, wherein exotic spin structures describe the Cooper
pairing phenomenon \cite{petry}. 
Multiply connected spaces are employed to study Feynman propagators \cite{banach1} as well as the vacuum polarization in electrodynamics \cite{ford}. Exotic spinors yield distinct  effects of vacuum polarization, producing a causal photon propagation.}  In the context of the exotic Dirac equation, or the Dirac-like type of coupled equations that govern Elko spinors, the electromagnetic potential is affected by the transformation $A \mapsto A + \frac{1}{2\pi i} \xi^{-1} \upd \xi$, namely, corresponding  to a shifted electromagnetic potential, with $\xi$ being a scalar field. When mass dimension 3/2 spinors (in 4D) or mass dimension 5/2 spinors (in 6D) are regarded, the exotic term  may be, hence, assimilated into any external electromagnetic potential, encoding an element of $H^1(M, \mathbb{Z}_2)$ \cite{petry,exotic, avis0}.  Hence, an additional term in the Dirac operator induces extra degrees of freedom for fermionic particles \cite{exotic}. In this case,  the 4D portion $D_{\mu}$ in \eqref{dcov} can be transformed into \cite{exotic,alex,isham0,petry,avis0}
\begin{gather}
 \gamma^M D_{M} \Upsilon(x^{\mu},r,\theta)=\left[\gamma^{M}\partial_{M}+ \gamma^{M}\Omega_{M}(r)+ \zeta_2\mathcal{F}_2(r)\right]\Upsilon(x^{\mu},r,\theta), 
\label{dcove}
\end{gather}
where the exotic term $\mathcal{F}_{2}$ reads 
\begin{equation}
\mathcal{F}_{2}(r)=\frac{1}{2\pi i}\xi^{-1}\upd {\xi}=\gamma^{M}\partial_{M}\Theta(r)\ ,
\label{exotic-gs}
\end{equation}
for a scalar field $\Theta(r)$ \cite{alex, exotic}. The exotic term in Eq. (\ref{exotic-gs})  originates from any non-trivial topology in the bulk \cite{alex, exotic} and has the restriction to be normalizable. It is worth to emphasize 
that only mass dimension one (in 4D) and mass dimension two (in 6D)  quantum fields are capable to probe exotic couplings. On the other hand, mass dimension five-halves (in 6D)  quantum fields realize this term as a shift of a gauge potential $A=\gamma^{M}A_{M}$. Therefore  any exotic term that might exist when gauge fields are considered in an effective Lagrangian is absorbed into a gauge potential $A'=\gamma^{M}(A_{M}+\partial_{M}\Theta(r))$  \cite{alex, exotic}. Hence, the topological bulk content, provided by the second cohomology group of the 6D space-time to $\mathbb{Z}_2$, 
generates a physical signature, corresponding to 
an additional term $\partial_{M}\Theta(r)$ in the covariant derivative. Nevertheless, standard fermions perceive it as a shift on a gauge potential \cite{exotic}.  
 Consequently just mass dimension one quantum fields can indeed be affected by any exotic coupling, since they do not interact with any gauge field.  Eqs. (\ref{dcove}) and (\ref{exotic-gs}) comprise a gauge choice, whose just the 4D portion is necessary 
 to circumvent the problems regarding imaginary values.

\begin{figure}[!htb]
       \centering
                \includegraphics[width=.68\linewidth]{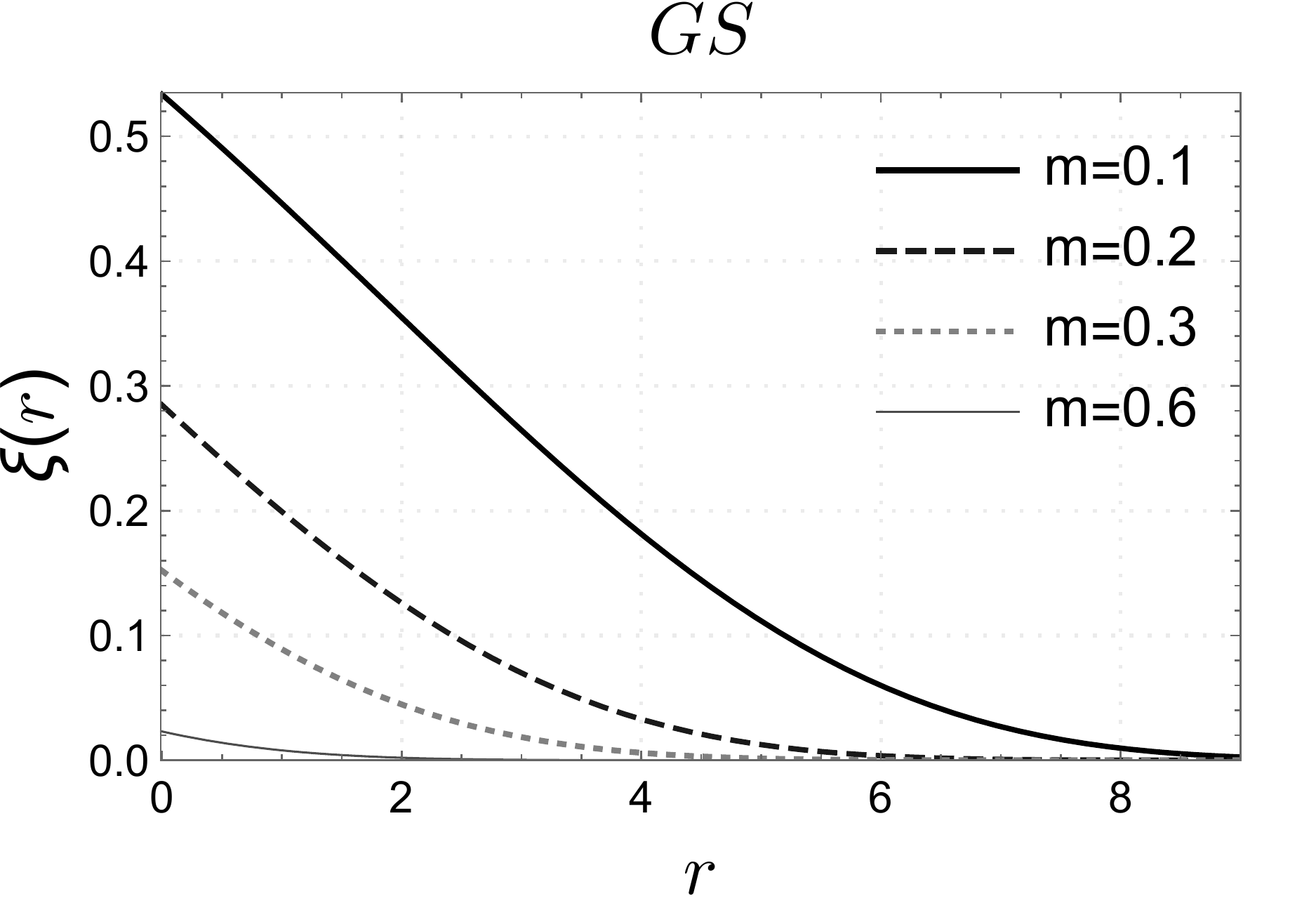}
                \caption{Scalar fields $\xi(r)$ for the GS thin model with $c=0.5$ and some mass values. All these $\xi(r)$ can be normalisable.}
                \label{GS-vf1}
\end{figure}

Similarly to proposed in  Ref. \cite{alex}, we can adjust the scalar fields $\xi$ in order to remove the complex-valued term in Eq. \eqref{sl-gs} as
\begin{gather}
\xi_{GS}(r)=\exp\left[-2\pi m \text{e}^{\frac{cr}{2}}\right] \ ,
\label{xi-gs}
\end{gather}
which has the plot shown in Fig. \ref{GS-vf1}.

Indeed, the field choice in Eq. \eqref{xi-gs} applied in Eq. \eqref{exotic-gs} shows  an exotic coupling $\mathcal{F}_2(r)=\frac{im_nc}{2}\ex^{\frac{cr}{2}} {dr}$. {Since that $dr \equiv \gamma^r$,} the term with coupling constant $\zeta_2=1$ cancels the imaginary term in Eq. \eqref{sl-gs}. Due to it, and with the previously fixation of Eq. \eqref{f-fix},  Eq. \eqref{sl-gs} turns to be:
\begin{gather}
\varepsilon_{n}''(r)-\frac{5c}{2}\varepsilon_{{n}}'(r)+m^2_n\ex^{cr}\varepsilon_{{n}}(r)=0.
\label{sl-gs-2}
 \end{gather}
which has only real-valued solutions in terms of Bessel functions \cite{GS1, Silva:2012yj, Diego,Davi3, Davi7} 
\begin{eqnarray}\label{bessel}
\varepsilon_n(r)=\ex^{\frac{5c}{4}r}\left\{N_1^n J_{\frac{5}{2}}\left(\frac{2m_n}{c}\ex^{\frac{cr}{2}}\right)+N_2^n Y_{\frac{5}{2}}\left(\frac{2m_n}{c}\ex^{\frac{cr}{2}}\right)\right\}\,.
 \end{eqnarray}
The oscillation of these solutions \eqref{bessel} exponentially grows  with $r$. Hence no massive mode can be trapped, as expected \cite{GS1, Silva:2012yj, Diego, Davi3, Davi7}. However, the resonance method can be applied in this case of Eq. \eqref{bessel}  \cite{Davi3, Davi7}, which it is not possible in the case of equation \eqref{m-gs}. Hereby, the main aim of this letter  has been achieved for the thin 6D model.

In order to study the generalisation of scalar coupling $\mathcal{F}_1$, we can perform the conformally plane metric transformation  ${\upd s_{6}}^2 = F(z) \left[\eta_{\mu\nu} \upd x^{\mu}\upd x^{\nu} + \upd z^{2} + \beta(z) \upd \theta^{2}\right]$, where $\beta(z) = H(z)/F(z)$. In this form, changing $z(r) =\int_{0}^{r}{F^{-\frac{1}{2}}(r')}\upd r'$ and $\varepsilon_{n}(z) = F^{-1}(z)\beta^{-\frac{1}{4}}(z)\tilde{\varepsilon}_n(z)$, assuming the fixation in Eq. \eqref{xi-gs}, we turn the Sturm-Liouville Eq. \eqref{sl} into a Schr\"{o}dinger-like equation as \cite{Silva:2012yj, Diego, Davi3, Davi7}:
\begin{equation}\label{schrodinger}
-\ddot{\tilde{\varepsilon}}_n(z) + U(z) \tilde{\varepsilon}_n(z) = m_n^2\tilde{\varepsilon}_n(z), \, \nonumber
\end{equation}
where the dots represent derivatives with respect to the $z$ coordinate and the analogue quantum potential $U(z)$ for the Elko has the form
\begin{gather}
\!\!\!\!\!\!U(z) \!=\! \frac{\ddot{F}}{F}\!+\!
\left(\frac{\dot{F}}{ 4F}\right)^2
\!-\!\frac{1}{8}\left(\frac{\dot{\beta}}{\beta}\right)^2\!+\!\frac{5}{8}\frac{\dot{F}}{F}\frac{\dot{\beta}}{\beta}\!+\!\frac{1}{4}\frac{\ddot{\beta}}{\beta}
\!+\!2\zeta_1\mathcal{F}_1 F \ . 
\label{Quantum-Potential}
\end{gather} 
On the other hand, the Ricci curvature in the conformal metric reads:
\begin{gather}
R(z)=\frac{1}{F}\left[5\frac{\ddot{F}}{F}+\frac{5}{2}\frac{\dot{F}}{F}\frac{\dot{\beta}}{\beta }+\frac{\ddot{\beta}}{\beta }-\frac{1}{2}\frac{\dot{\beta}^2}{\beta^2}\right].
\label{ricci-z}
\end{gather}

Hence, we conclude in this new variable system that, unlike 5D models, a scalar coupling proportional to Ricci curvature cannot be used for a general 6D model. In fact, the general expression for the $\mathcal{F}_1$ fields, that allows the existence of Elko zero mode, must  be:
\begin{gather}
\zeta_1\mathcal{F}_1(z)=-\frac{1}{32 F}\left[5 \left(\frac{\dot{F}}{F}\right)^2+\frac{\dot{\beta}}{\beta}\frac{\dot{F}}{F}+\left(\frac{\dot{\beta}}{\beta}\right)^2\right]
\label{fz-fix}
\end{gather}
which has none second derivative terms, as the present in the Ricci scalar \eqref{ricci-z}. However, for the particular GS thin model this coupling can be interpreted as  the curvature, due to the non-regular term $\beta(z)= constant$, which clearly vanishes when derived. Nevertheless, in this new set of variable, the Schr\"{o}dinger potential and the normalised zero mode in the GS models reads
\begin{gather}
U(z)=6\left(z+\frac{2}{c}\right)^{-2}\ , \   \tilde{\varepsilon}_0(z)=\sqrt{\frac{24}{c^3}}\left(z+\frac{2}{c}\right)^{-2}\,,
\label{uez}
\end{gather}
allowing the confinement of massless Elko in the thin 6D model.

\section{Elko in the thick HC model} In the Hamilton string-cigar thick model, the factors \eqref{hc-string} replaced into coefficients \eqref{pq} of differential equation \eqref{sl} read
\begin{gather}\label{pq-hc}
\mathcal{P}(r)=-\frac{5c}{2}\left(\tanh^2{(cr)}-\frac{4}{5}{\sech(2cr)}\right),\\
\mathcal{Q}_n(r)=\frac{m^2_n}{\ex^{-(cr+\tanh{cr})}}
-\frac{im_nc\tanh^2{(cr)}}{2\ex^{-\frac{1}{2}(cr+\tanh{cr})}}-2\zeta_1\mathcal{F}_1+\nonumber\\-\frac{5c^2}{16}\left[\tanh^4(cr)+\frac{4}{5}\left({2\text{sech}^2(2cr)}-\frac{\sech^2(cr)}{\coth(cr)} \right ) \right ].
\end{gather}
The analytical solutions of the equation \eqref{sl}  with coefficients \eqref{pq-hc} are  not straightforward to find, even for the massless case. However, we see that asymptotically when $r\to 0$, the coefficients \eqref{pq-hc} exhibit  singularities, whereas for   $r\to\infty$, the coefficients \eqref{pq-hc} converge those thin coefficients of Eq. \eqref{sl-gs}, $\mathcal{P}(r\to \infty)=-\frac{5c}{2}$ and $\mathcal{Q}_n(r\to \infty)=m^2_n\ex^{cr}-\frac{im_nc}{2}\ex^{\frac{cr}{2}}-\frac{5c^2}{16}- 2 \zeta_1\mathcal{F}_1$. Consequently, the massive eigenfunctions for the HC string are divergent at infinity, hence non-normalizable, irrespectively of the boundary condition at the origin.

\begin{figure}[!htb]
 \centering
 \includegraphics[width=.68\linewidth]{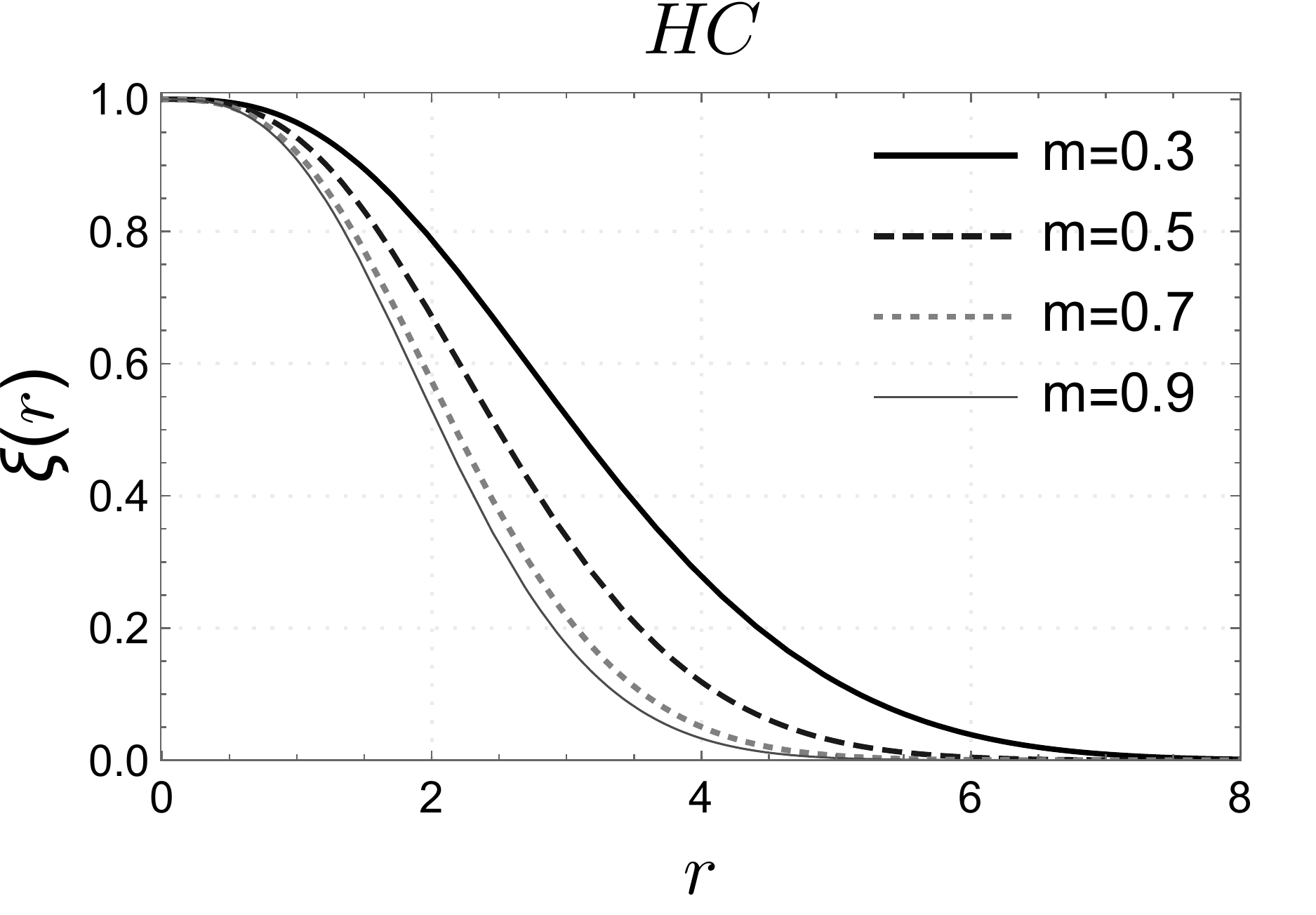}
 \caption{Numerical plot of scalar fields $\xi(r)$ for HC thick mode with $c=0.5$ and some mass values. All these $\xi(r)$ can be normalisable.}
                \label{HC-vf2}
\end{figure}

For the zero mode in HC, we can propose a generalised coupling into $r$ variable in this form:
\begin{gather}
\zeta_1\mathcal{F}_1(r)=-\frac{1}{2}\left[\left(\frac{F'}{2F}\right)^2+\left(\frac{H'}{4H}\right)^2\right]\,,
\nonumber \\
=-\frac{c^2}{8}\left[\tanh ^4(c r)\!+\!\frac{1}{4} \left[4 \text{csch}(2c r)\!-\!\tanh ^2(c r)\right]^2\right]\,,
\label{f-fix-hc}
\end{gather}
which can be put in the alternative $z$ variable form of Eq. \eqref{fz-fix} when $\frac{\upd}{\upd r}\mapsto F^{-\frac{1}{2}}\frac{\upd}{\upd z}$  and $H=\beta F$. This scalar solution of Eq. \eqref{f-fix-hc} is not the Ricci curvature in the Eq. \eqref{sricci}. However, the solution in Eq. \eqref{f-fix-hc} has asymptomatically the same behaviour of the string-vortex scalar field of Ref. \cite{gio}. This imposition of Eq. \eqref{f-fix-hc} in Eq. \eqref{sl} gives a generalised constant zero mode that obey the conditions \eqref{normb} and \eqref{normc} as $
\varepsilon_0(r)=\left[\int_0^{\infty} dr F(r)\sqrt{H(r)}\right]^{-{1}/{2}}. 
\label{hc-0}$ 
This expression can be only numerically evaluated. Moreover, in order to remove the imaginary mass term, we can fix the scalar fields $\zeta_2\mathcal{F}_2(r)=\frac{1}{2\pi i}\frac{\upd \xi}{\xi}=\frac{im}{2}F'(r)F^{-\frac{3}{2}}(r) {dr}$ and $\xi(r)$ as:
\begin{gather}
\xi(r)=\exp{\left[\pi m\int_{r}\upd r{F'F^{-\frac{3}{2}}}\right]}. 
\end{gather}
Fig. \eqref{HC-vf2} shows that this scalar field $\xi(r)$ is regular and normalisable in the HC model, as well.

\begin{figure}[!htb]
       \centering
\includegraphics[width=.68\linewidth]{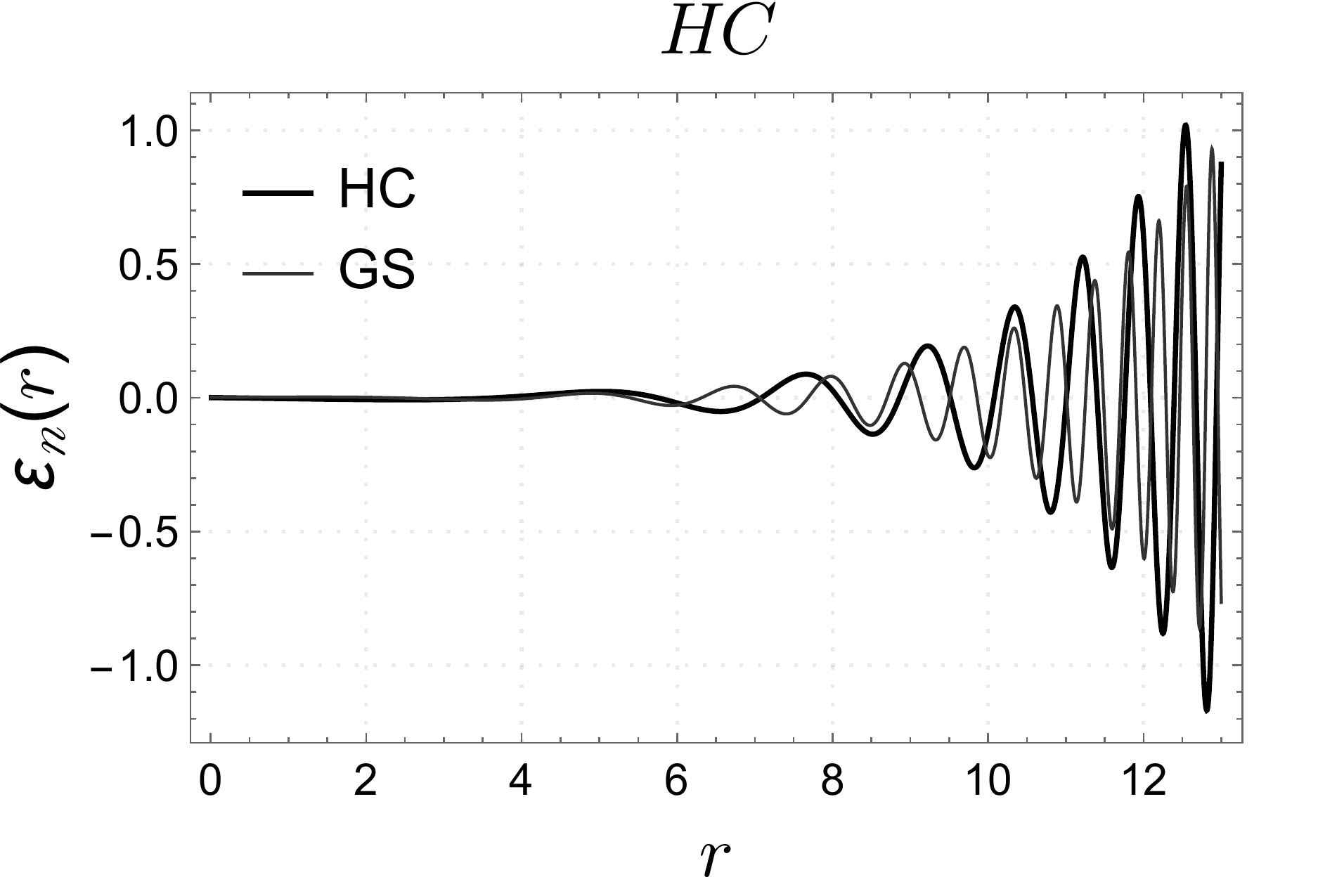}
\caption{Numerical solution $\varepsilon_n(r)$ for the HC thick model \eqref{sl-hc-2} (black line) and for GS thin model \eqref{bessel} (gray line), both with $c=0.5$ and $m=0.8$. The normalisation constants  are set $N_1^n=N_2^n=1/1500$ for this interval.}
\label{HC-mass2}
\end{figure}

\begin{figure}[!htb]
\centering
\includegraphics[width=.68\linewidth]{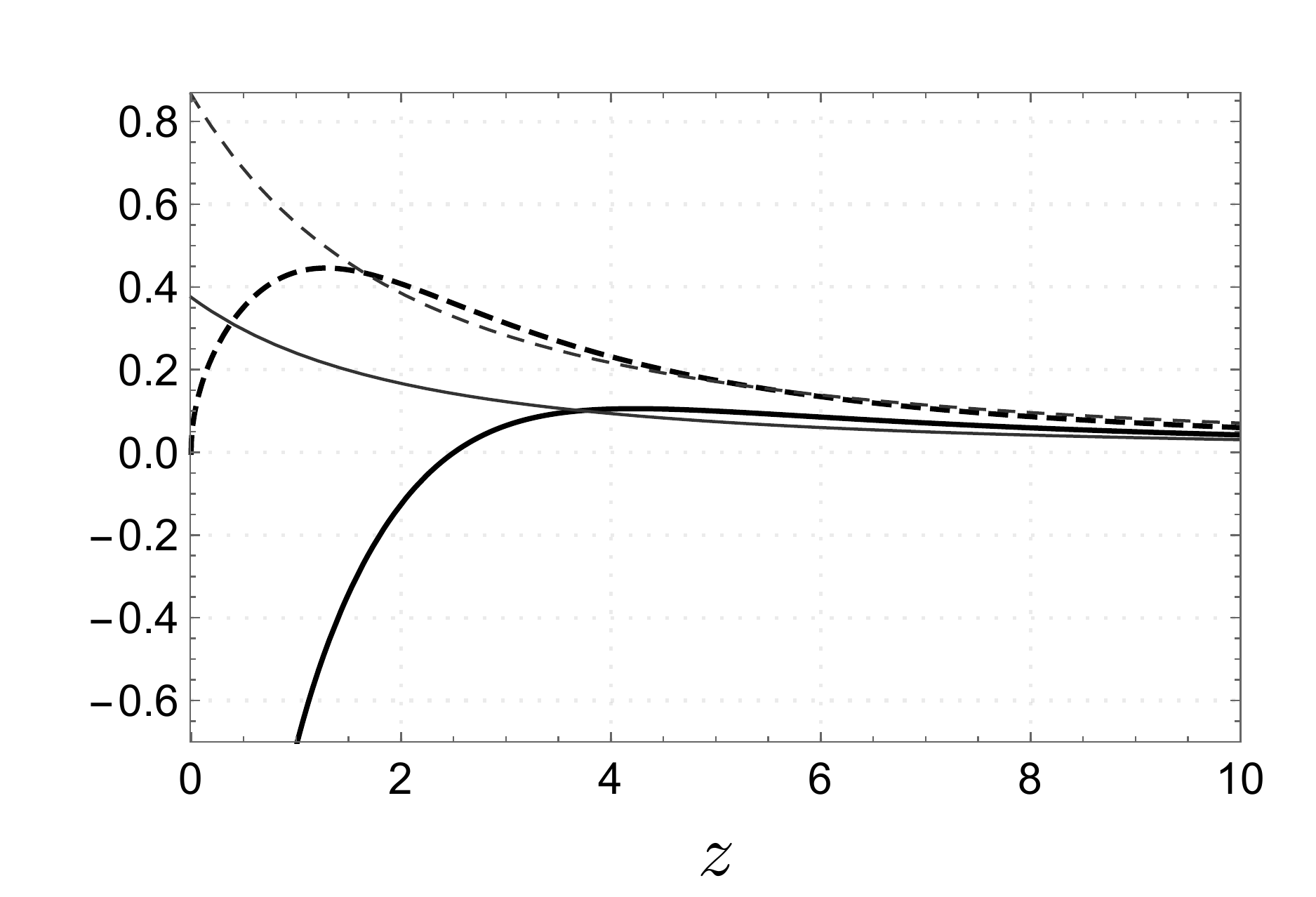}
\caption{Plots for potentials $U(z)$ (full lines) for HC thick model (black line) and for GS thin model (gray line), and normalised massless modes $\varepsilon_0(z)$ (dashed lines) for HC thick model (dashed black line) and for GS thin model (dashed gray line),  all  cases for $c=0.5$.}
\label{u-0}
\end{figure}

Hence, the real-valued Sturm-Liouville equation \eqref{sl} turns into the HC model as:
\begin{gather}
\varepsilon_{n}''(r)-\frac{5c}{2}\left(\tanh^2{(cr)}-\frac{4}{5}{\sech(2cr)}\right)\varepsilon_{{n}}'(r)+\nonumber \\+m^2_n\ex^{(cr-\tanh{cr})}
\varepsilon_{{n}}(r)=0.
\label{sl-hc-2}
 \end{gather}
this solution was studied in Ref. \cite{Diego}. The numerical value differs from the analytical solution  of the thin model \eqref{bessel} only close to the origin \cite{Diego}, as can be see in Fig. \ref{HC-mass2}. 
Working with the conformal variable $z$, the expressions for the Elko zero mode, $\varepsilon_0(z)$, and analogue quantum potential $U(z)$ can be only numerically-valued in the HC scenario. However, Fig. \ref{u-0} shows a comparasion between the results for the GS model \eqref{uez} and those for the HC model. In the HC model, the zero mode is also normalisable, but the HC potential has an infinite well at origin, as also verified in Refs. \cite{Diego, Davi3}. We will study the resonances in a future work. 

\section{Conclusions}
 
 We have proved that issues of Elko spinor fields in 5D can be solved in 6D brane-worlds. We analyse the Elko fields in a thin and a thick string-like scenarios in 6D, where one uses  two type of scalar fields to confine the zero mode, turning  the massive modes real-valued. The results in the GS thin model have been  analytically achieved, whereas in the HC model the results were only numerically valued. Finally, we generalised the result for all string-like metrics in 6D.  For future works, we propose the generation of a 6D model by a topological abelian Higgs model, where the background exotic term can be naturally obtained from this geometry. The study of other types of coupling is also present among our aims. Moreover, further mass-dimension one fields (in 4D) or mass dimension 5/2 fields (in 6D), like the self-interacting ones for any spin \cite{chengk} and singular spinor fields  \cite{Villalobos:2015xca}, are going to be analysed in a 6D model context, accordingly. It is also worth to point out that given that Elko is a prime dark matter candidate, the associated KK modes should be localized  to the brane \emph{a posteriori}. However, it might live \emph{a priori} in the bulk, since we have shown that the zero mode can be trapped on the brane. It would explain why dark matter only weakly interacts with matter.

\acknowledgments
CASA thanks CNPq  for financial support through grants 305766/2012-0 and 448142/2014-7. DMD thanks to Projeto CNPq UFC-UFABC  304721/2014-0 and CAPES. RdR is grateful to CNPq grant No. 303293/2015-2, and to FAPESP Grant No. 2015/10270-0.

\end{document}